\documentclass[review,authoryear]{elsarticle}

\usepackage{graphicx}
\usepackage[utf8]{inputenc}
\usepackage{url}
\usepackage{tabularx}
\usepackage{longtable}
\usepackage{listings}
\usepackage{natbib}
\usepackage{tabulary}
\usepackage{longtable}
\usepackage{amsmath}
\usepackage{hyperref}

\newcommand{\superscript}[1]{\ensuremath{^{\textrm{#1}}}}

\journal{Journal of Informetrics}

\begin{document}

\begin{frontmatter}

\title{Visualization of Co-Readership Patterns from an Online Reference Management System}

\author[pk]{Peter Kraker\corref{cor1}}
\ead{pkraker@know-center.at}
\author[cs]{Christian Schlögl}
\ead{christian.schloegl@uni-graz.at}
\author[kj]{Kris Jack}
\ead{kris.jack@mendeley.com}
\author[sl]{Stefanie Lindstaedt}
\ead{slind@know-center.at}

\cortext[cor1]{Corresponding author}
\address[pk]{Know-Center, Inffeldgasse 13, 8010 Graz, Austria}
\address[cs]{University of Graz, Universitätsstraße 15, 8010 Graz, Austria}
\address[kj]{Mendeley, 144a Clerkenwell Road, EC2 RF London, UK}
\address[sl]{Know-Center, Inffeldgasse 13, 8010 Graz, Austria}

\begin{abstract}
In this paper, we analyze the adequacy and applicability of readership statistics recorded in social reference management systems for creating knowledge domain visualizations. First, we investigate the distribution of subject areas in user libraries of educational technology researchers on Mendeley. The results show that around 69\% of the publications in an average user library can be attributed to a single subject area. Then, we use co-readership patterns to map the field of educational technology. The resulting visualization prototype, based on the most read publications in this field on Mendeley, reveals 13 topic areas of educational technology research. The visualization is a recent representation of the field: 80\% of the publications included were published within ten years of data collection. The characteristics of the readers, however, introduce certain biases to the visualization. Knowledge domain visualizations based on readership statistics are therefore multifaceted and timely, but it is important that the characteristics of the underlying sample are made transparent.
\end{abstract}


\end{frontmatter}

\section{Introduction}

In recent scientometric literature, usage data is being discussed as a valuable alternative to citations. With the advent of e-journals, digital libraries, and web-based archives, click and download data have been suggested as a potential alternative to citations \cite[]{Kurtz2005, Rowlands2007}. Compared to citation data, usage data has the advantage of being available earlier, shortly after a paper has been published. In many instances, usage statistics are also easier to obtain and collect \cite[]{Bollen2005, Brody2006, Haustein2011}. Furthermore, usage statistics allow for an analysis of publications and research outputs  that do not receive citations or for which citations are not tracked \cite[]{Priem2010}.

Another type of usage data besides clicks and downloads is created in social reference management systems like BibSonomy\footnote{\url{http://bibsonomy.org}} and Mendeley\footnote{\url{http://mendeley.com}}. These systems enable users to store their references in a personal library and share them with other people. The number of times an article has been added to user libraries is commonly referred to as the number of readers, or in short readership\footnote{Initially, the term readership might seem a bit misleading, because the addition of an article to a user library does not guarantee that the article has actually been read by said user. Nevertheless, researchers need to make a second decision after downloading an article before they add it to their user libraries. Furthermore, the term is already well established among researchers (see e.g. \cite{Bar-Ilan2012,Haustein2014,Thelwall2014,Zahedi2014}); thus we use it in our research for reasons of consistency and to avoid neologisms.}.

Readership statistics have been of high scientometric interest in recent years. It has been shown that readership statistics provide a good coverage of top publications~\cite[]{Bar-Ilan2012}, and that there is a medium correlation between readership data and citations~\cite[]{Schloegl2013} and a medium to high correlation between the impact factor and journal readership~\cite[]{Kraker2012a}. Furthermore, \cite{Jiang2011} employ readership statistics from CiteULike to form clusters based on the occurrence and co-occurrence of articles in user libraries. They also correlate these clusters with ISI subject categories, and find them as effective as citation-based clusters when removing journals that cannot be found in CiteULike. 

\begin{figure}
  \includegraphics[width=0.5\textwidth]{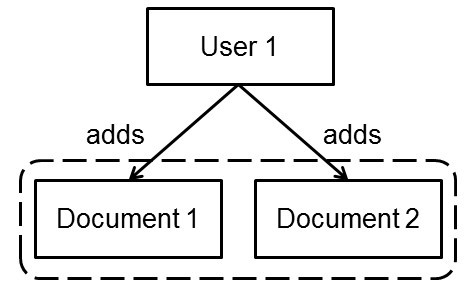}
	\caption{Co-readership of two documents is established when at least one user has added the two documents to his or her user library.}
	\label{fig:coreadership}
\end{figure}

\begin{figure}
\includegraphics[width=0.75\textwidth]{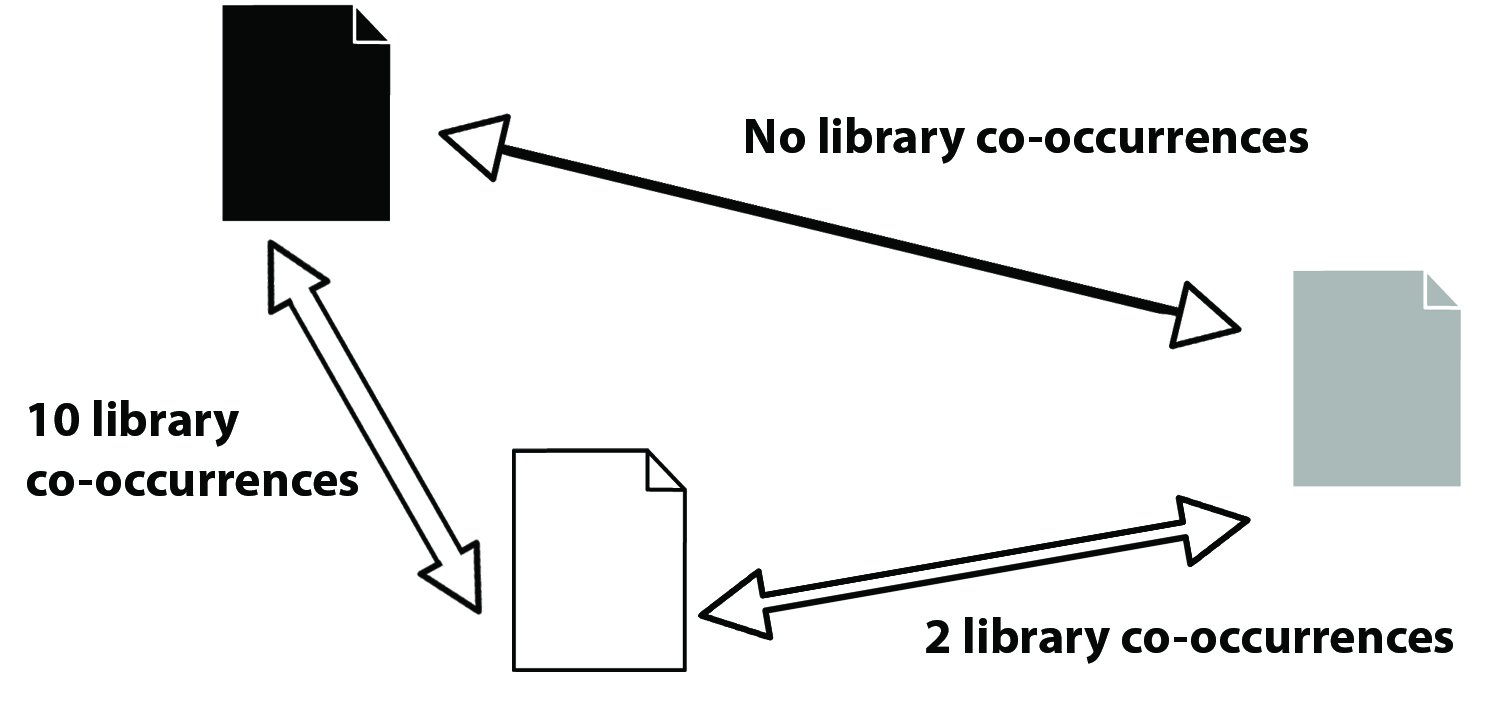}
\caption{Relationships between documents in a field based on co-readership. Co-occurrence in user libraries is employed as a measure of subject similarity.}
\label{fig:readership}
\end{figure}

Therefore, we consider co-readership as a measure of subject similarity. Co-readership relation between two documents is established when at least one user has added the two documents to his or her user library (see Figure~\ref{fig:coreadership}). We assume that the more often the same two documents have been added to user libraries, the more likely they are of the same or a similar subject. The topical relationship established by co-readership can then be exploited for visualizations by clustering those papers that have high co-readership numbers (see Figure~\ref{fig:readership}). To the best of our knowledge, this measure has not been exploited before for knowledge domain visualization.

In this study, we first investigate the distribution of subject areas in user libraries of educational technology researchers on Mendeley. Then, we employ co-readership patterns for knowledge domain visualization to explore the field of educational technology. Educational technology is multi-disciplinary and highly dynamic in nature, as it is influenced by changes in pedagogical concepts and emerging technologies~\cite[]{Siemens2009}, as well as social change~\cite[]{Czerniewicz2010}. Therefore, it seemed to be especially appropriate for this kind of analysis.

\section{Related Work}

\begin{figure}
  \includegraphics[width=\textwidth]{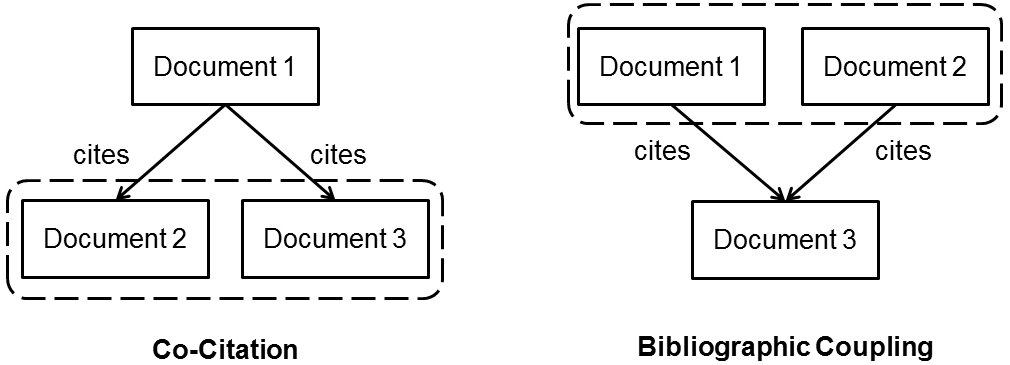}
	\caption{Relationships between documents on the basis of citations~\citep[adapted from][]{Schlogl2001}}
	\label{fig:citation_relationships}
\end{figure}

Traditionally, knowledge domain visualizations are based on citations. \cite{Small1973} and \cite{Marshakova1973} proposed co-citation as a measure of subject similarity and co-occurrence of ideas (see Figure~\ref{fig:citation_relationships}, left side, for a graphical representation of the relationship). This relationship can be employed to cluster documents, authors, or journals from a certain field and to map them in a two-dimensional space. Co-citation analysis has been used to map many fields, for instance information management \cite[p. 48]{Schlogl2001}, hypertext \cite[]{Chen1999}, and educational technology~\cite[]{Chen2011} to name just a few. Furthermore, co-citation analysis has also been used to map out all of science \cite[]{Small1999, Boyack2005}.

There is, however, a significant problem with citations: they take a long time to appear. It takes around two to six years after an article is published before the citation count peaks \cite[]{Amin2003}. Therefore, visualizations based on co-citations - and indeed all analyses that are based on incoming citations - have to deal with this time lag. Bibliographic coupling \cite[]{Kessler1963} presents an alternative to co-citation analysis; it is formed when two documents cite the same source document (see Figure~\ref{fig:citation_relationships}, right side). The more publications in the reference list the two documents have in common, the more related they are.

Bibliographic coupling is based on outgoing citations available at the time of publication and can therefore be used to map the research front. One difference between bibliographic coupling and co-citation analysis is that the former is a retrospective method \cite[]{Garfield2001}, which means that the relationship between two documents cannot change over time.
%Another downside of bibliographic coupling is that prior knowledge is needed to define which publications are part of the research front. 
For an overview of the properties and the accuracy of the two citation-based mapping techniques refer to \citet[chap. III.4]{Egghe1990} and \cite{Boyack2010}.

In contrast to citations, usage statistics have been almost exclusively used in evaluative scientometrics~\citep[see e.g.][]{Darmoni2002, Bollen2007, Schloegl2010}. There are only a handful of examples in relational scientometrics and knowledge domain visualization. One of the first are \cite{Polanco2006}, who propose to use co-occurrences of document requests for clustering and mapping. \cite{Bollen2006} use consecutive accesses to journal articles as a measure of journal relationships. They derive clusters of journals which are statistically significantly related to ISI subject categories. In a later study, \cite{Bollen2009a} create an overview map of all of science. The authors collect hundreds of millions of user interactions with digital libraries and bibliographic databases. Then, they re-create click-streams for each user, aggregated by journal, and apply network analysis to them. Among the challenges of the approach, the authors name that clickstreams need to be aggregated from various data sources. The varying user interfaces and the difference between reader and author population may introduce biases to the visualization~\cite[]{Bollen2008b}. 

In social reference management systems we can address these challenges. First, we are able to use library co-occurrence from a single service as a basis for mapping the intellectual structure of a scientific domain. Second, being able to precisely attribute papers to individual readers allows for a better understanding of the results as the information found in the user profile 
%(geographic location, career stage, research interests, biographical information) 
adds further context. With the help of profile information, we can for example analyze the influence of different user groups. When using library co-occurrence, however, we are missing the temporal aspect represented in clickstreams, which may play a role when establishing subject similarity.

\section{Data Source}

All data for this study was sourced from Mendeley on 10 August 2012. Mendeley provides users with software tools that support them in conducting research~\cite[]{Henning2008}. One of the most popular of these tools is Mendeley Desktop, a cross-platform, freely downloadable PDF and reference management application. It allows users to organize their personal libraries into folders and apply tags to them for later retrieval. The articles, added by users around the world, are then crowd-sourced into a single collection called the Mendeley research catalog~\cite[]{Hammerton2012}. At the time of writing, this catalog contains more than one hundred million unique articles, crowd-sourced from over two and a half million users.

The users of Mendeley do not only help with building the catalog but also with structuring it. Users can identify themselves as belonging to a scientific discipline and optionally also to a sub-discipline. In August 2012, Mendeley offered 25 disciplines (see Table~\ref{tab:disciplines}), and 473 sub-disciplines (see Table~\ref{tab:sub_disciplines} for the sub-disciplines of ``Education''). Each time, a user from a certain (sub-)discipline adds a document to his or her library, the document is automatically assigned to this (sub-)discipline in the catalog\footnote{As a result, a document can be assigned to more than one (sub-)disciplines.}.

Furthermore, Mendeley Web enables users to create and maintain a user profile that includes their discipline, organization, location, career stage, research interests, biographical information, education, professional experience, contact details, and their own publications.

\begin{table}%
	\begin{tabulary}{\textwidth}{|L|L|}
		\hline
		Arts and Literature & Astronomy / Astrophysics / Space Science \\ 
		\hline
		Biological Sciences & Business Administration \\ 
		\hline
		Chemistry & Computer and Information Science \\ 
		\hline
		Design & Earth Sciences \\ 
		\hline
		Economics & Education \\ 
		\hline
		Electrical and Electronic Engineering & Engineering \\ 
		\hline
		Environmental Sciences & Humanities \\ 
		\hline
		Law & Linguistics \\ 
		\hline
		Management Science / Operations Research & Materials Science \\ 
		\hline
		Mathematics & Medicine \\ 
		\hline
		Philosophy & Physics \\ 
		\hline
		Psychology & Social Sciences \\ 
		\hline
		Sports and Recreation &  \\ 
		\hline
	\end{tabulary}
\caption{List of the 25 disciplines in the Mendeley catalog (Source: \protect\url{http://www.mendeley.com/research-papers/})}
\label{tab:disciplines}
\end{table}

\begin{table}
	\begin{tabulary}{\textwidth}{|L|L|}
		\hline
		Business Education & Comparative Education  \\ 
		\hline
		Counselling & Curriculum Studies \\ 
		\hline
		Education Research & Educational Administration  \\ 
		\hline
		Educational Change & Educational Technology \\ 
		\hline
		Language Education & Mathematics Education \\ 
		\hline
		Medical Education & Miscellaneous  \\ 
		\hline
		Physical Education & Science Education \\ 
		\hline
		Sociology of Education & Special Education \\ 
		\hline
		Teacher Education & Testing and Evaluation \\ 
		\hline
	\end{tabulary}
\caption{List of the 18 sub-disciplines of ``Education'' (Source: \protect\url{http://www.mendeley.com/disciplines/education/})}
\label{tab:sub_disciplines}
\end{table}

The following data sets have been sourced on 10 August 2012 and represent data for the sub-discipline educational technology that had been accumulated in the system up to that point:

\begin{itemize}
	\item User profiles and user libraries: all user profiles and their accompanying user libraries in the sub-discipline of educational technology (n=2,154 users)\footnote{User profiles and user libraries were sourced at a later point (23 January 2013). Only users that signed up before 10 August 2012 were considered to ensure congruency with the rest of the data set. However, (minor) shifts in the user base cannot be excluded, since in the case users changed their sub-discipline, Mendeley provided only the most recent one chosen.}
	%\item User libraries: all user libraries in the sub-discipline of educational technology (n=1,677 user libraries)\footnote{The difference between users (2,154) and user libraries (1,677) is explained by the fact that not all users on Mendeley keep a library. Some use it only as a social networking tool; others have signed up once but never added any documents.}
	\item Documents: metadata of all documents in the field of educational technology (n=144,500 documents)
	\item Co-occurrences: co-occurrences of these documents in all Mendeley user libraries (n=56,049,431 co-occurrences).\footnote{Co-occurrence calculation is a computationally intensive process. Therefore, the number of documents per user library was limited to 500. If a user library contained more documents, 500 documents were randomly selected. Then the co-occurrences were calculated.}
\end{itemize}

\section{Distribution of Subject Areas in User Libraries}

Subject homogeneity is a necessary precondition that the results of co-readership analysis are valid; otherwise the assumption that co-occurrence of articles in user libraries implies subject similarity cannot be upheld. Therefore, we analyzed the subject distribution of articles included in Mendeley user libraries and compared it to the subject area distribution of reference lists of articles in Web of Science. The basis of this analysis is the user profiles and user libraries data set of researchers in educational technology (n=2,154 users). As already mentioned, categorization of users into sub-disciplines is determined by self-ascription of users.

In a first step, we analyzed the distribution of journal articles in user libraries. We used SCImago, which is a bibliometric service based on the bibliographic database Scopus, as an external validation source. SCImago categorizes each journal into one of 28 subject areas. The documents from the field of educational technology were matched to these subject areas through the journals they appear in. We used a semi-automated approach for matching journal names in Mendeley and SCImago\footnote{Journal names from both sources were transliterated (if necessary) and converted to lowercase. White space at the beginning and at the end was stripped. Colons, commas, and dashes were removed as well as a potential starting definite article “The”. The resulting strings were compared, and all complete matches were taken. In a next step, the list of matches was searched for near-misses and other apparent mismatches, e.g. “User Model-ling and User-Adapted Interaction” as compared to “User Modelling and User-Adapted Interactions”}. After this procedure, 1,107 user libraries, which contained at least one article in a journal that is indexed by SCImago, were left.

\begin{figure}
\centering
\includegraphics[width=\textwidth]{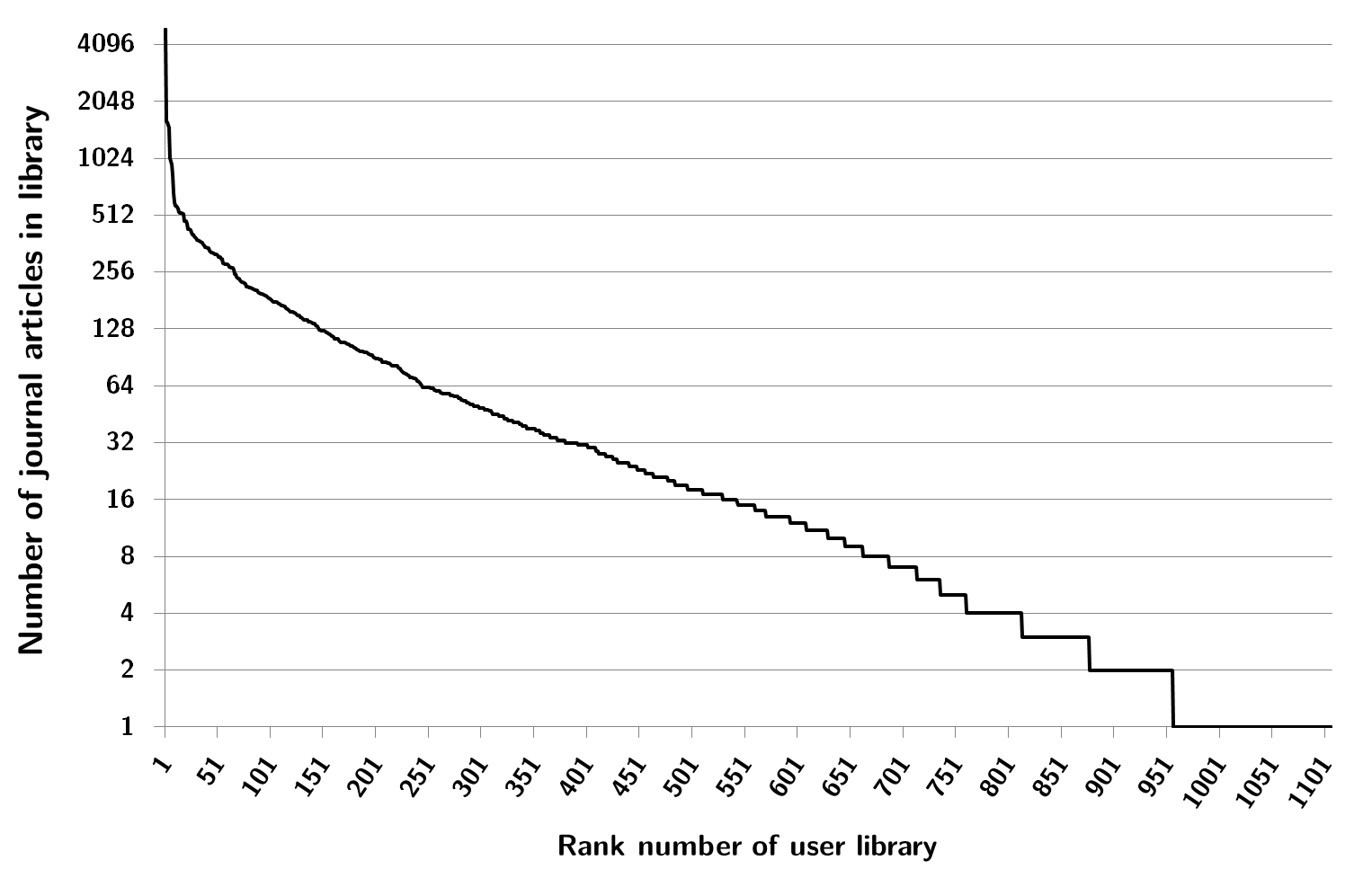}
\caption{Distribution of the size of user libraries (no. of journal articles) in educational technology on a logarithmic scale (n=72,721 journal articles in 1,107 user libraries)}
\label{fig:user_libraries}
\end{figure} 

A user library in educational technology has on average 155.7 documents (SD=460, Median=17); slightly more than a third (56.7) of these documents are on average journal articles that appeared in journals indexed by SCImago (SD=202.2, Median=15). As Figure~\ref{fig:user_libraries} shows, the distribution of the size of user libraries (number of journal articles) is highly skewed; 
%In 151 libraries, there is only one journal article, whereas there are only 17 libraries that contain 500 or more journal articles. 
10\% of all user libraries cover 62.4\% of total journal articles.

\begin{figure}
\centering
\includegraphics[width=\textwidth]{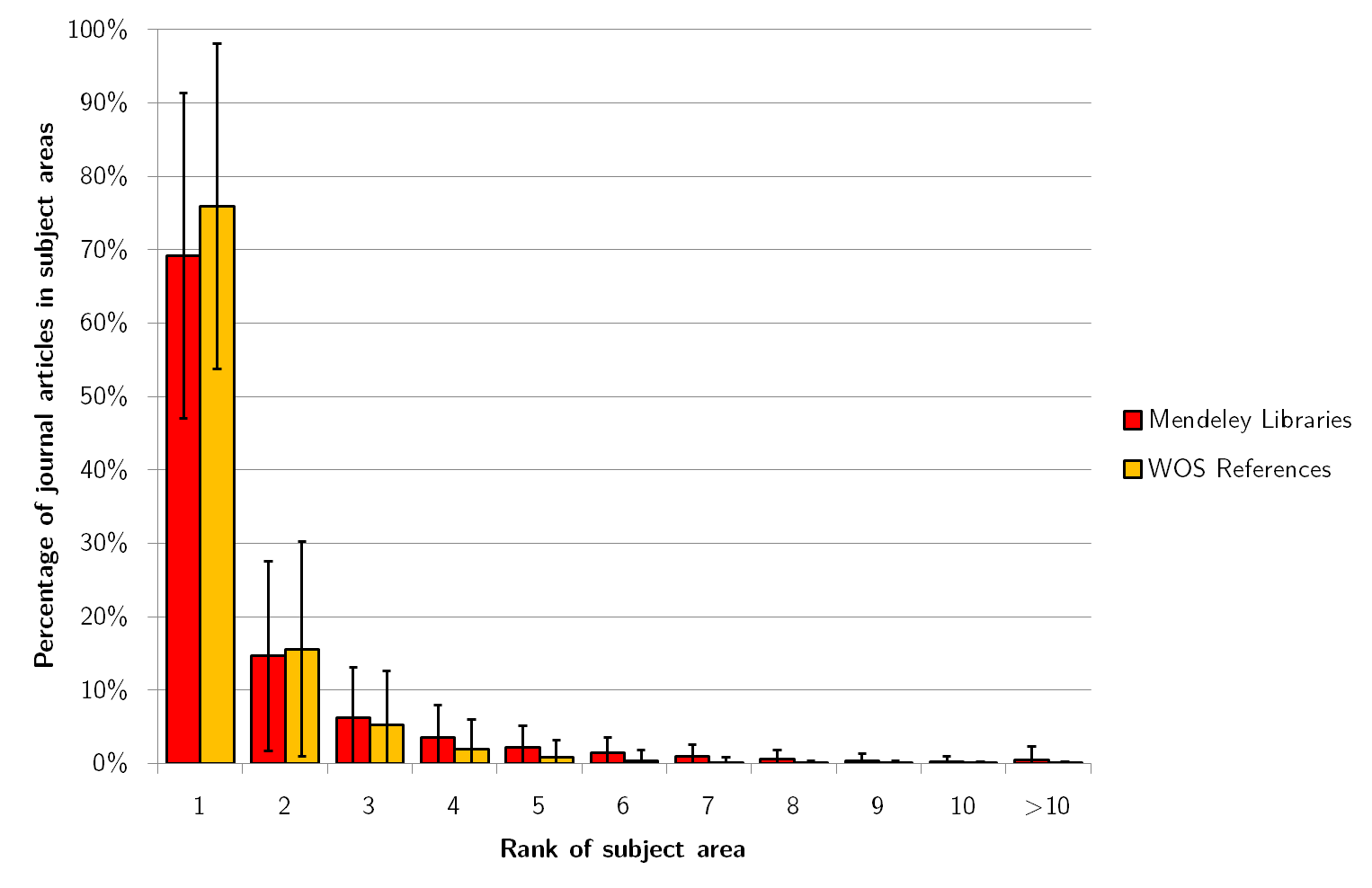}
\caption{Subject area frequency distribution of articles in user libraries from educational technology (n=72,721 journal articles in 1,107 user libraries) and cited references in WoS articles (n=13,841 cited references in 1,394 documents). Ranks 11-25 (Mendeley) and 11-12 (WOS articles) were summed up.}
\label{fig:subject_areas}
\end{figure}

We also created a data set of cited references from Web of Science. We searched for articles and reviews with the topic ``educational technology'' in the WOS Core Collection. This resulted in 1,394 documents. We retrieved the cited references for these documents; each document has on average 29.2 cited references (SD=23.8, Median=25). We then applied the procedure outlined above to match references to subject areas via their journals. This resulted in 1221 reference lists which contained at least one document that is indexed by SCImago; 38\% of these (11.1 documents) are on average journal articles that appeared in journals indexed by SCImago (SD=12.7, Median=7).

%Furthermore, we also analyzed the subject area distribution of all 49,970 unique articles contained in these user libraries (see Figure~\ref{fig:distribution_unique_articles}). 53.6\% of all articles can be attributed to social science. Psychology follows second with 10.5\%. In total, 27 subject area were present in the data, 20 of which have been subsumed under ``Other`''. Note that educational technology is a multi-disciplinary field with three major contributing disciplines: education, computer science, and psychology. There are also secondary contributions by library and information science, broadcasting and mass communication, curriculum development, evaluation, and sociology \cite{Ely2008}.
%
%\begin{figure}
%\centering
%\includegraphics[width=\textwidth]{distribution_unique_articles.png}
%\caption{Distribution of SCImago categories among unique articles in user libraries of educational technology users (n=49,970 articles).}
%\label{fig:distribution_unique_articles}
%\end{figure}

Finally, we calculated the distribution of SCImago categories for each Mendeley user library from educational technology and each cited reference list for the article set retrieved from Web of Science. Afterwards, we ranked the results by subject area. For each library, the percentage of articles that are categorized into a common subject area was calculated. Then, the areas were ranked according to their frequency. The average subject area distribution for all educational technology user libraries can be seen in Figure~\ref{fig:subject_areas}.

\begin{table}
	\begin{tabularx}{\textwidth}{l|lllll}
		\hline
		 & \multicolumn{2}{c}{Mendeley} &  & \multicolumn{2}{c}{Web of Science} \\ 
		\cline{2-6}
		Subject Area & \multicolumn{1}{c}{Mean} & \multicolumn{1}{c}{SD} &  & \multicolumn{1}{c}{Mean} & \multicolumn{1}{c}{SD} \\ 
		\hline
		1 & 69.19\% & 22.18\% &  & 75.91\% & 22.12\% \\ 
		2 & 14.65\% & 12.94\% &  & 15.59\% & 14.63\% \\ 
		3 & 6.23\% & 6.83\% &  & 5.24\% & 7.38\% \\ 
		4 & 3.59\% & 4.42\% &  & 1.93\% & 4.06\% \\ 
		5 & 2.14\% & 2.95\% &  & 0.80\% & 2.37\% \\ 
		6 & 1.41\% & 2.15\% &  & 0.35\% & 1.44\% \\ 
		7 & 0.97\% & 1.65\% &  & 0.11\% & 0.70\% \\ 
		8 & 0.61\% & 1.17\% &  & 0.03\% & 0.33\% \\ 
		9 & 0.41\% & 0.87\% &  & 0.02\% & 0.28\% \\ 
		10 & 0.29\% & 0.67\% &  & 0.01\% & 0.20\% \\ 
		>10 & 0.51\% & 1.77\% &  & 0.01\% & 0.17\% \\ 
		\hline
	\end{tabularx}

	\caption{Subject area frequency distribution of articles in user libraries from educational technology (n=72,721 journal articles in 1,107 user libraries) and cited references in WoS articles (n=13,841 cited references in 1,394 documents). Ranks 11-25 (Mendeley) and 11-12 (WOS articles) were summed up.}
	\label{tab:subject_area_distribution}
\end{table}

For Mendeley, on average, 69.2\% of articles in a user library fall into the top subject area (see Table~\ref{tab:subject_area_distribution}). 14.6\% of articles in libraries were on average assigned to the second most frequent second area, while only 6.3\% and 3.6\% of articles are devoted to the third and fourth ranked subject area respectively. For WoS, 76.0\% of cited references in journal articles fall into the top subject area. 15.6\% of articles were assigned to the second most frequent area, and 5.2\% into the third. In Mendeley, three subject categories account for more than 90\% of all articles in an average user library, whereas in journal articles, the top two categories account for more than 90\% of all cited references. 

These results show that, as was expected, cited references in journal articles are very homogeneous with regards to their subject area distribution. Mendeley user libraries are less homogeneous, and they spread out over more subject areas. The top subject area, however, still accounts for 69.2\% of articles in an average user libraries (compared to 76.0\% in cited references), even though the number of journal articles in an average user library (56.7) is 5 times higher than the number of cited references in an average journal article (11.2). Therefore, although co-readership probably offers a weaker indication of subject similarity than co-citation, it can still be expected to serve as a useful indication of subject similarity.
%We therefore draw the conclusion that the assumption that co-readership establishes subject similarity can be upheld, as long as appropriate thresholds are set. 

%In 79.7\% of cases (882 out of 1,107 libraries) the top subject category was social science. The other 225 libraries are spread over 22 top subject categories, with medicine being the top category in 3.9\%, psychology in 3.4\%, arts in 2.6\%, and computer science in 2.4\% of libraries.

\section{Visualization of Co-Readership Patterns}

\begin{figure}
\centering
\includegraphics[width=\textwidth]{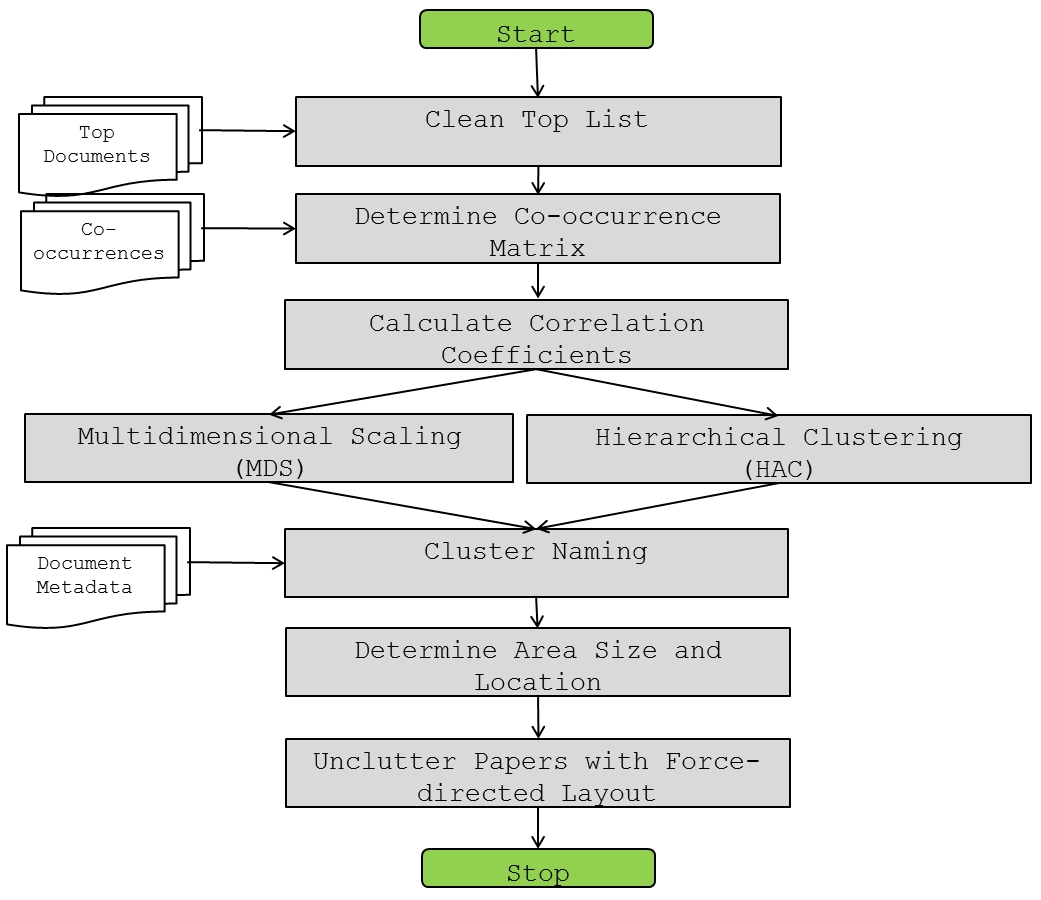}
\caption{Overview of the procedure used to create the co-readership visualization.}
\label{fig:procedure}
\end{figure}

For the visualization of co-readership patterns, we followed the knowledge domain visualization process as proposed by~\cite{Borner2003}. It consists of four steps: (1) selection of an appropriate data source, (2) determination of the unit of analysis, (3) analysis of the data using dimensionality reduction techniques, and (4) visualization and interaction design. Each of these steps is detailed below.
The whole procedure can be seen in Figure~\ref{fig:procedure}.

\subsection{Data Selection and Pre-processing}
The documents included in the analysis were taken from the Mendeley sub-discipline of educational technology\footnote{\url{http://www.mendeley.com/disciplines/education/educational-technology/}}. As mentioned before, a document is added to a sub-discipline, if it has at least one reader from this sub-discipline. At the point of data collection, there were approximately 2,150 users that had indicated educational technology in their user profile. 

To retrieve the most important documents, the document list was sorted by the number of library occurrences within the sub-discipline. A threshold of 16 occurrences was introduced as selection criterion. This means, a document needs to have been added to at least 16 libraries owned by users who identified themselves as being in the field of educational technology to be included in the analysis, leading to a total of 91 documents. The threshold was chosen upon manual inspection of the results. Among the evaluated solutions (thresholds between 11 and 25), the solution with 16 occurrences had the highest purity (0.91)\footnote{Purity is an external cluster evaluation technique. It is defined as the number of correctly assigned documents divided by the number of all documents. A document is correctly assigned when it corresponds to the class that is most frequent within its cluster\cite[p.356f]{Manning2009}}.
%For smaller threshold values (and consequently, larger amounts of documents), the resulting solutions had lower purity values. 
Since sub-discipline is an optional field in Mendeley, only a minority of users have filled out this field. In order to include more users in Mendeley, the co-occurrence calculation was extended to all user libraries. The 91 documents appeared in 7,414 user libraries with a total of 19,402 co-occurrences.

In a next step, a co-occurrence matrix was created. In line with \cite{Mccain1990}, diagonal values were treated as missing values. In addition, document pairs with no combined readership were treated as missing values.\footnote{Usually, these cases are put down as zero co-occurrences. As mentioned above, however, we were limited to a maximum of 500 documents per library when calculating the co-occurrences due to computational constraints. Therefore, we cannot say for sure whether no co-occurrence was found and thus we put down these cases as missing values. We did not find much difference between the two variations, but in the case of missing values, topics were not spread over clusters and the solution was more stable in a bootstrapping analysis. One reason for this could be that the matrix in co-readership analysis is less sparse than in co-citation analysis. Treating document pairs with no combined readership as missing values might therefore serve as a better indicator of discrimination between documents. Therefore, the missing values approach was chosen. Nevertheless, it remains to be determined whether this will hold true for future data sets.}

Based on the co-occurrence matrix, we computed the Pearson correlation coefficient matrix with pairwise complete observations. These correlation coefficients were then used to calculate Euclidean distances between the documents.\footnote{Note that the Pearson correlation coefficient is disputed as a measure of subject similarity~\cite[]{Ahlgren2003}. For a discussion of alternate similarity measures see e.g. \cite{Egghe2010}.}

\subsection{Clustering and Mapping}

The matrix of correlation coefficients was the basis for multidimensional scaling (MDS) and hierarchical agglomerative clustering (HAC). Multidimensional scaling was used to project the documents into a two-dimensional space, clustering to find topic areas in the projection. For hierarchical agglomerative clustering, we employed Ward's method (minimum variance) using the R command \textit{hclust}. Ward's method successively merges those two clusters that minimize the increase in the total within-cluster variance~\cite[p. 510]{Hair2010}. It is known to join smaller clusters and to produce clusters of approximately the same size~\cite[p. 523]{Tan2007}. 

The number of clusters was determined by the elbow method using the R function \textit{elbow.batch}. This function defines an elbow when the number of clusters k explains at least 80\% of the variance in the model, and when the increment is lower than 1\% for a bigger k. This criterion was reached at an explained variance of 84\% and lead to a total of 13 clusters.

In a second step, we plotted the results in a two-dimensional space with non-metric multidimensional scaling (NMDS). NMDS is often employed in scientific mapping efforts. Examples can be found in \cite{White1998} and in \cite{Tsay2003}. NMDS is an iterative approach: beginning with a random start configuration, it tries to minimize a given stress function in consecutive steps. Since NMDS is prone to reaching local minima, usually a number of random starts are used to find an optimum solution. 

We selected the NMDS implementation provided by the R ecodist package~\cite[]{Goslee2007}. 
%It uses a modified stress function:
%
%\begin{equation}
	%Stress = \sqrt{\sum^{}_{h,i}(d_{hi}-\hat{d}_{hi})^2/\sum^{}_{h,i}\hat{d}^2_{hi}}
	%\label{eq:modified_stress}
%\end{equation}
%\\
%$d_{hi}$: \text{dissimilarity between samples h and i}\\
%$\hat{d}_{hi}$: \text{distance predicted by regression}\\
%This implementation produced more clearly separable clusters in a visual sense in comparison to implementations that use the original stress function proposed by~\cite{Kruskal1964} while maintaining a similar stress value and a similar R\superscript{2}. 
In the NMDS, stress is reported as 0.2 and the R\superscript{2} is reported as 0.86. According to~\cite{Hair2010}, acceptable results for R\superscript{2} start at 0.60.

To create labels for the clusters, titles and abstracts of the documents in each cluster were submitted to the APIs of Zemanta\footnote{\url{http://zemanta.com}} and OpenCalais\footnote{\url{http://opencalais.com}}. Both services crawl the semantic web and return a number of concepts that describe the content. The returned concepts were compared to word n-grams generated from titles and abstracts. The more words a concept has (and therefore, the more information it contains), and the more often it occurs within the text, the more likely it is to be the label of the cluster. The results of this procedure were manually checked and corrected if needed.

\begin{figure}[!ht]
  \includegraphics[width=\textwidth]{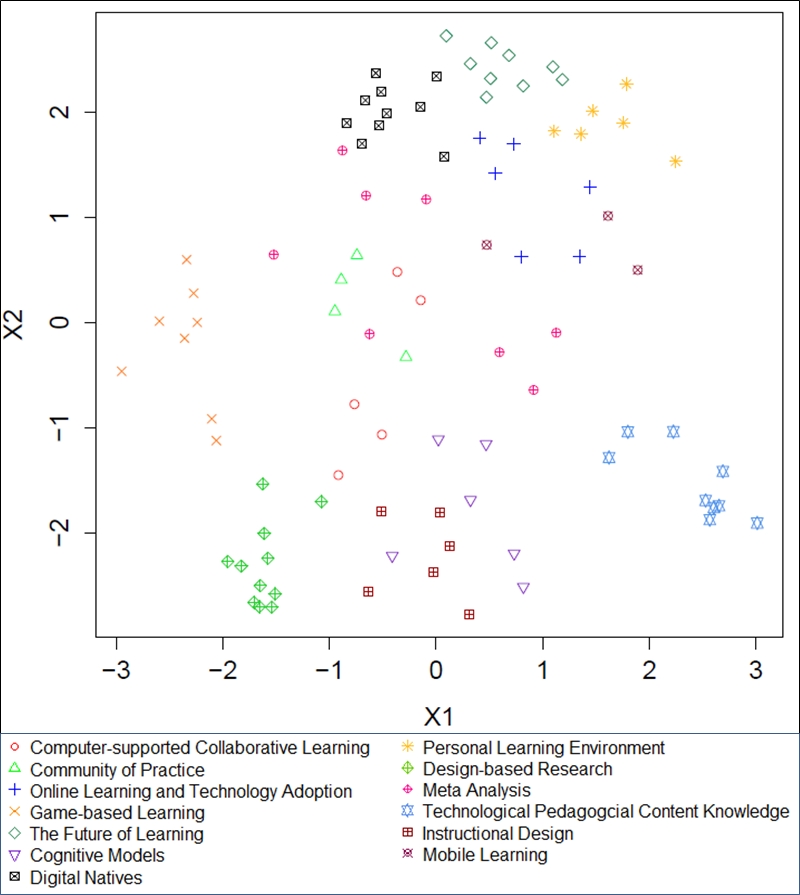}
\caption{Result of NMDS, HAC, and the naming algorithm. Each symbol represents a document, all of the documents with the same symbol constitute a topic area.}
\label{fig:mds_clustering_naming}      
\end{figure}

A plot of the results from the procedure described above can be seen in Figure~\ref{fig:mds_clustering_naming}. Each symbol represents a document. The type of symbol signifies the topic area it belongs to. These 13 areas are listed in the legend below the graph. 

\subsection{Web Visualization}

In order to allow users to interact with this graph, we developed an interactive web visualization prototype. The visualization was created using D3.js\footnote{\url{http://d3js.org}}. In the prototype, documents are represented as rectangles with dogears, a common metaphor, used in many icons and graphics. The size of the document signifies the number of readers it has. To avoid coding the documents with symbols (as in Figure~\ref{fig:mds_clustering_naming}), topic areas are represented as bubbles. The center of each bubble is calculated as the mean of the coordinates of the publications based on the NMDS result. The size of the bubble is determined by the number of combined readers of the publications in the topic area.

Additionally, force-directed placement~\cite[]{Fruchterman1991} was employed on the documents to unclutter the visualization and move documents into their respective topic areas\footnote{The area centers were denoted as gravitational centers. Documents not within the limits of the topic area were instructed to move towards the gravitational center. Edges and corresponding edge weights were not set; they are therefore initialized to default values by D3.}. To prevent overlapping documents, the collision detection algorithm by Mike Bostock\footnote{\url{http://bl.ocks.org/mbostock/3231298}} was used.

It is important to note that - in contrast to the topic areas - the relative closeness of documents in the visualization does not necessarily imply relative subject similarity.\footnote{In the uncluttering effort using force-directed placement, the positions of documents are changed in a way that does not necessarily preserve the relative distances. Therefore, the distances between documents in the visualization do not represent the distances calculated with MDS anymore.} To review the relationship between individual papers, one needs to go back to the original output of the MDS shown in Figure~\ref{fig:mds_clustering_naming}.

\subsection{Results}

\begin{figure}[!ht]
\includegraphics[width=\textwidth]{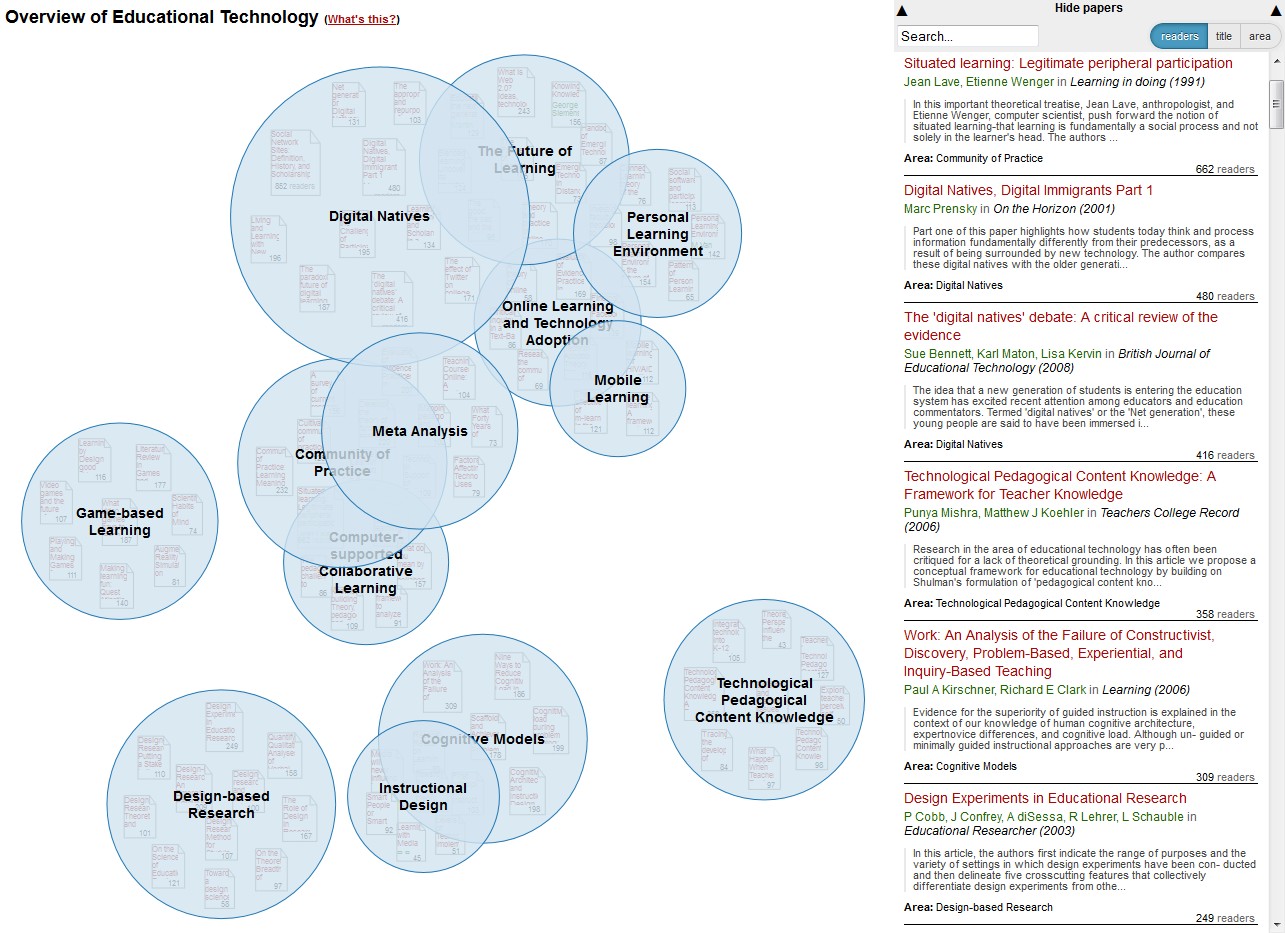}
\caption{Knowledge domain visualization of educational technology. The bubbles represent topic areas within the domain. The size of a bubble relates to the number of combined readers.}
\label{fig:headstart1}
\end{figure}

The resulting visualization prototype, which can be accessed on Mendeley Labs\footnote{\url{http://labs.mendeley.com/headstart}. The source code can be obtained from \url{https://knowminer.at/svn/opensource/other-licenses/lgpl_v3/headstart/}}, is shown in Figures~\ref{fig:headstart1}. In the first few seconds of the visualization, the force-directed placement algorithm is executed. The papers are untangled and pulled into their respective areas, represented by the blue bubbles. After the force-directed algorithm has finished, users can interact with the visualization. 

%Regarding the interaction design, we followed the well-tested approach of ``overview first, zoom and filter, then details-on-demand''~\cite[]{Shneiderman1996}. Once a user clicks on a bubble, he or she is presented with relevant documents for that area (see Figure~\ref{fig:headstart2}). The meta data of each document is displayed in the document representation itself. It consists of the most common meta data: title, author(s), year, and journal/conference name (if applicable). 
%
%\begin{figure}[!ht]
  %\includegraphics[width=0.6\textwidth]{headstart_final2.jpg}
	%\caption{Zooming into the area ''Technological Pedagogical Content Knowledge''}
  %\label{fig:headstart2}
%\end{figure}
%
%The dropdown on the right displays the same data in list form with an added abstract. By clicking on one of the documents, a user can access all meta data for that document. If a preview is available, it can be retrieved by clicking on the thumbnail in the meta data panel. In addition, a user can filter the publications by entering terms in the search field on top of the list (see Figures~\ref{fig:headstart1} and \ref{fig:headstart2}). Only publications that contain all of the search terms (Boolean AND) are displayed within the bubbles and the list. The list can be sorted by title, area, and number of readers to facilitate exploration via the list format.

\subsubsection{Topic Area Description and Distribution}

As can be seen in Figure~\ref{fig:headstart1}, there are 13 topic areas in the visualization with a combined readership of 13,630 at the time of data collection (10 August 2012). Table~\ref{tab:areas} gives an overview of the topic areas. It shows that they differ in terms of the number of documents and the number of readers. \textit{Digital Natives} has the highest readership with over 20\% of all readers. It has twice the readership of the second largest area: \textit{Design-based Research} (DBR). DBR includes the most documents (11) of all areas. \textit{Community of Practice} has only four documents, but still sports the fourth most readers. The area with the least readers and the least number of documents is \textit{Mobile Learning} with just 3 documents and a combined readership of 345. 

%The map is mostly topical, with two exceptions: \textit{Meta Analysis} is a collection of reviews/state-of-the-art analyses, and \textit{Design-based Research} represents a specific method. \textit{The Future of Learning} is also somewhat orthogonal as it describes technological developments. 

The topic areas can again be assigned to meta-areas. These meta-areas are formed by areas that are close to each other, as is assumed by multidimensional scaling.  On the top of the map (see Figure~\ref{fig:headstart1}), social and technological developments are being discussed (in \textit{Digital Natives} and \textit{The Future of Learning}). Beneath, there is a large cluster of learning methods and technologies, spanning \textit{Mobile Learning}, \textit{Personal Learning Environment}, \textit{Online Learning and Technology Adoption}, \textit{Community of Practice}, and \textit{Game-based Learning}. On the bottom of the visualization, there is a cluster of topic areas that form the psychological, pedagogical, and methodological foundations of the field. The areas \textit{Computer-supported Collaborative Learning}, \textit{Instructional Design} and \textit{Cognition} relate to psychology, while \textit{Technological Pedagogical Content Knowledge} relates to pedagogy. Research methods are represented by \textit{Design-based Research}. 

From what was mentioned above, it follows that pedagogical and psychological topics are covered very well in the visualization. However, topic areas that are largely influenced by computer science such as \textit{Adaptive Hypermedia} or knowledge management (e.g. \textit{Work-integrated Learning}) are missing from the overview. The reason for this is most likely the discipline taxonomy in Mendeley (see section~\ref{subsec:biases}).

%Another characteristic of multidimensional scaling is that it shows central and peripheral areas due to their placement on the map \cite[]{Mccain1990}. Right in the middle, the area \textit{Meta Analysis} contains reviews of the field. Its central position stems most likely from the fact that these reviews relate to many of the surrounding areas (cp. Figure~\ref{fig:mds_clustering_naming}), and that they appear in many user libraries together because of their comprehensive nature. Other central areas are \textit{Computer-supported Collaborative Learning}, \textit{Communities of Practice}, \textit{Online Learning and Technology Adoption}, and \textit{Online Learning and Technology Adoption}. The remaining areas are more peripheral in the visualization; \textit{Game-based Learning}, \textit{Design-based Research}, and \textit{Technological Pedagogical Content Knowledge} in particular are placed further to the edges of the visualization.

\begin{table}
\begin{tabularx}{\textwidth}{X|r|r|r}
Topic Area & No. Documents & No. Readers & \% Readership \\ 
\hline
Digital Natives & 10 & 2,865 & 21.0\% \\ 
\hline
Design-based Research & 11 & 1,477 & 10.8\% \\ 
\hline
The Future of Learning & 9 & 1,183 & 8.7\% \\ 
\hline
Community of Practice & 4 & 1,175 & 8.6\% \\ 
\hline
Cognitive Models & 6 & 1,169 & 8.6\% \\ 
\hline
Technological Pedagogical Content Knowledge & 9 & 1049 & 7.7\% \\ 
\hline
Game-based Learning & 8 & 993 & 7.3\% \\ 
\hline
Meta Analysis & 8 & 991 & 7.3\% \\ 
\hline
Personal Learning Environment & 6 & 648 & 4.8\% \\ 
\hline
Online Learning and Technology Adoption & 6 & 637 & 4.7\% \\ 
\hline
Computer-supported Collaborative Learning & 5 & 615 & 4.5\% \\ 
\hline
Instructional Design & 6 & 483 & 3.5\% \\ 
\hline
Mobile Learning & 3 & 345 & 2.5\% \\ 
\hline
Sum & 91 & 13,630 & 100.0\% \\ 
\end{tabularx}
\caption{Topic areas in the visualization}
\label{tab:areas}
\end{table}

\subsubsection{Publication Types and Age of Publications}

The 91 documents in the visualization can be divided into five different types of publications. The majority are journal articles (71 items, or 78\%), followed by reports (7), books (6) and book chapters (5), and conference papers (2). The 71 journal articles were published in a variety of journals. The highest number of articles was published in ``Computers \& Education'' (8), followed by ``Educational Technology Research \& Development'' and ``The Internet and Higher Education'' (both 6) and ``Review of Educational Research'', ``Educational Researcher'' and ``Educational Psychologist'' (all 5). These publication outlets are among the highest impact journals in the Journal Citation Reports \cite[]{ThomsonReuters2013}. All of the documents in the visualization are in English.

\begin{figure}
  \includegraphics[width=\textwidth]{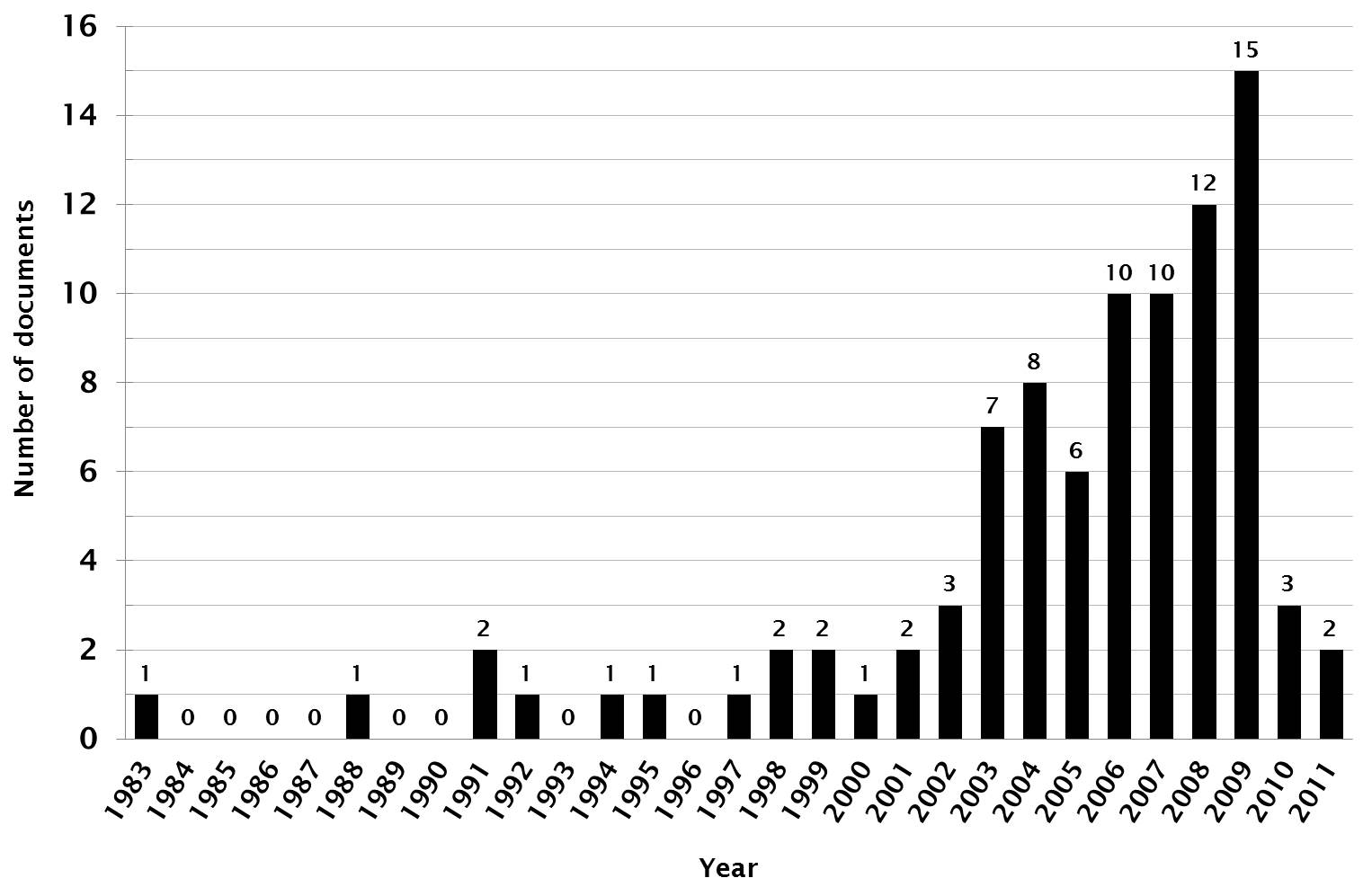}
\caption{Distribution of publication years of documents in the visualization (n=91)}
\label{fig:year_distr}
\end{figure}

Figure~\ref{fig:year_distr} shows the age distribution of the 91 publications covered in the visualization. 80\% of publications were published from 2003 onwards, meaning that they were younger than ten years at the time of data collection (10 August 2012). Most documents were published in 2009. The median age of publications is 6.0 years (Mean = 7.3 years)\footnote{Calculation based on the exact date of data collection on 10/08/2012}. The relative small amount of publications from 2010 and 2011 can be explained by the circumstance that it is more difficult for recent publications to reach the threshold value than for older ones.

%Classics within the field are still contained in this visualization; for the most part they inform research that is still prevalent today. Examples are ``Situated learning: Legitimate peripheral action''~\cite[]{Lave1991} or ``Cognitive load during problem solving: Effects on learning'' \cite[]{Sweller1988}. An exception is the area ``Instructional Design'' which contains only documents that were published before 2003. Here, the classic media debate between Clark and Kozma is represented, as well as other older papers relating to instructional design.

\section{Discussion}

\subsection{Recency}

In the conducted co-readership analysis, the mean age of publications is 7.3 years with 80\% of articles published within 10 years of data collection. While this constitutes a contemporary selection of publications, the relative low proportion of articles younger than two years indicates that not all of the latest developments might be represented in the visualization. However, in a comparable co-citation mapping effort in educational technology by \cite{Cho2012}, the mean age of papers was 14.1 years (Median = 14 years) which is almost double the age of publications in the co-readership analysis. In addition, only 18\% of the 28 papers included in the co-citation analysis were less than 10 years of age\footnote{All calculations based on the publication year of the most recent article in the analysis (2011)}. 

This suggests that the results of a co-readership analysis may be much more up-to-date than co-citation analysis. In contrast to bibliographic coupling, however, there is still a certain time lag after publication that needs to be taken into account. Therefore, a co-readership analysis may be most appropriate when a contemporary overview is desired but a dynamic method is preferred over a static one. 

%In comparison to bibliographic coupling, however, co-readership visualization has a couple of advantages; first of all it is a dynamic method, meaning that the results can change over time. Second, the data employed (readership statistics) allow to select the publications to be included in the analysis by the information given in the user profile.

\subsection{Biases in the Visualization}\label{subsec:biases}
An analysis of the results shows that the visualization is not free from biases. First, all of the papers are in English, even though educational technology is often researched by local communities that communicate in their native language \cite[]{Ely2008}. Second, the knowledge domain visualization represents an education-dominated view that lacks topic areas related to computer science.

Biases in usage statistics analyses were first mentioned by \cite{Bollen2008} in a study of downloads in an institutional repository. The authors found great differences in the correlation of usage impact factor and journal impact factor depending on the user base. The authors therefore concluded that these biases occur due to sample characteristics. 

Consequently, we looked into the sample characteristics of users in educational technology that we investigated based on their user profiles (n=2,153 user profiles). At first, we analysed the geographical distribution of users. One of the reasons for the fact that all of the papers are in English is surely that English is the \textit{lingua franca} in science and research~\cite[]{Tardy2004}. But most likely, this dominance of English also stems from the fact that there is a strong bias towards English-speaking countries on Mendeley.

\begin{figure}[!ht]
  \includegraphics[width=\textwidth]{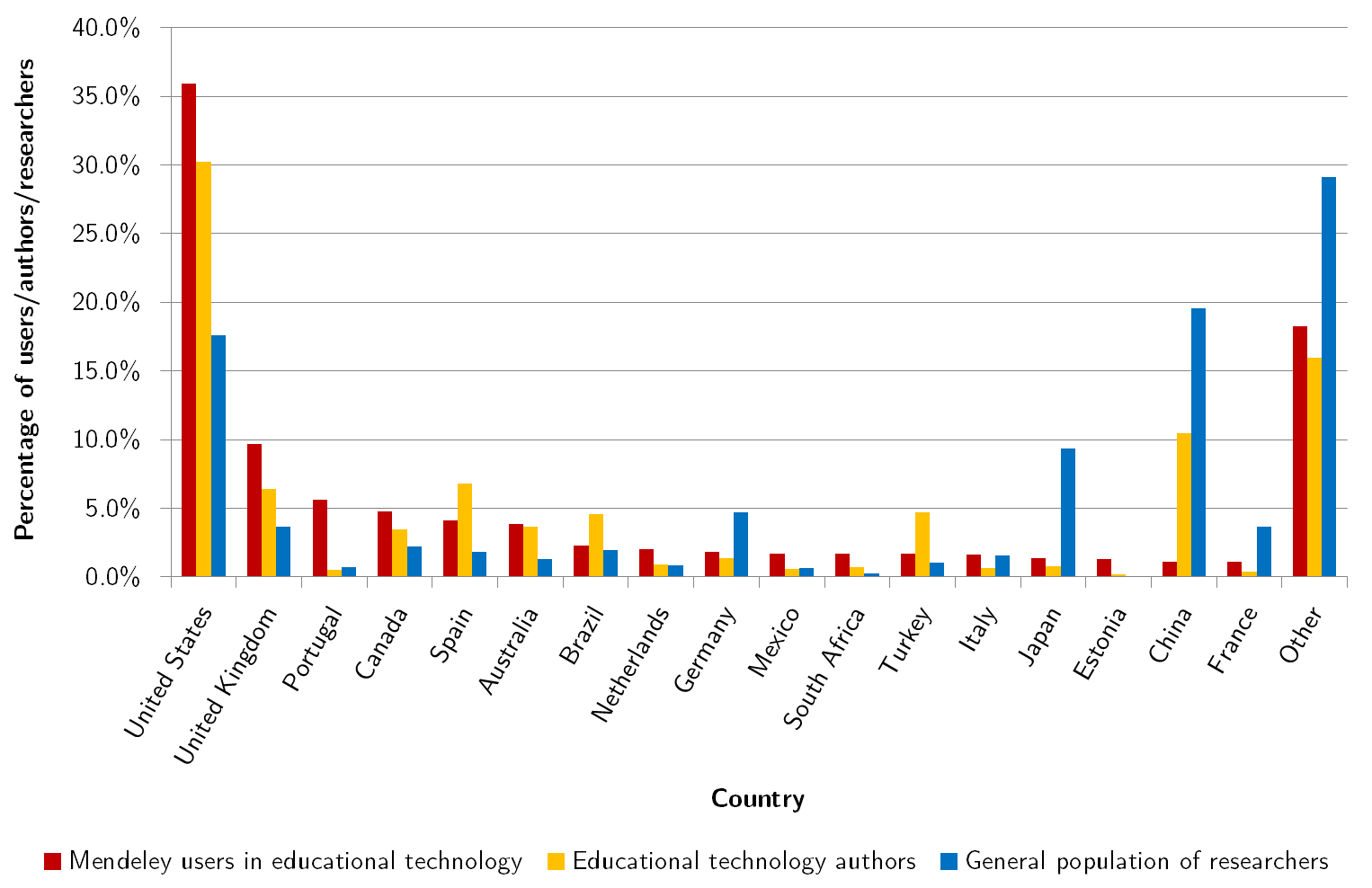}
\caption{Geographic distribution of users from educational technology in Mendeley (n=927 users), educational technology authors in WoS (n=2,965 authors) and researchers in general (n=7,043,655 researchers). Only countries that were present in all three data sets were taken into account when calculating the distribution. Data sources: Mendeley/Web of Science Core Collection/World Bank Development Indicators (Researchers in R\&D (per million people); Population, total)}
\label{fig:user_distribution_comparison} 
\end{figure}

This assumption is backed up by the results of the geographical analysis (see Figure~\ref{fig:user_distribution_comparison}). Out of 2,153 users, 927 (43.1\%) have chosen to list a country in their user profile. In total, 70 countries have been specified, but the distribution is highly skewed. There is an emphasis on the US and the UK with a total number of 423 users (45.6\%). In fact, when adding Canada and Australia, English-speaking countries have a share of over 54.3\%. 56 countries with a low share of users have been summed up under ``Other'' (21.7\%). This shows that Mendeley users come from a wide variety of countries, but that there is a strong focus on English speaking countries.

Figure~\ref{fig:user_distribution_comparison} shows the comparison of this distribution to the geographic distribution of educational technology authors in the Web of Science Core Collection and the distribution of researchers according to \cite{TheWorldBank2014}. Of the 2,965 unique authors with an assigned geographic location that have contributed to an article with the topic ``educational technology'' (out of 4,602 in total), 1,298 (43.8\%) come from one of the four major English-speaking countries. Although this proportion is very high (for instance, the share of researchers from these 4 countries is 24.8\% according to the World Bank), the dominance of Mendeley users from these 4 countries is even stronger. 
%In particular, users from China and Japan are only weakly represented in Mendeley. Conversely, the share of authors is larger than the share of users in Spain, Brazil, Turkey, and especially China (1.1\% in Mendeley vs. 19.6\% in WOS). In the geographic distribution of researchers in general, the English-speaking countries only have a share of 28.1\%. China and Japan have a combined share of 32.8\% of researchers, whereas they only account for 2.5\% of readers in the Mendeley category of educational technology.
Two facts play an important role with regards to this imbalance: first, Mendeley originated in the UK and has an office in the USA. Second, the Mendeley software is only available in English for now.

The bias towards disciplines strongly related to education can be explained by Mendeley's discipline taxonomy which was used to determine the paper pre-selection in this study. Even though educational technology is an interdisciplinary field, it appears solely as a sub-discipline of education. The sign-up process in Mendeley requires a user to first select a discipline such as education, social science, or computer and information science. In a second step, a user can select a sub-discipline, such as educational technology. Therefore, a scholar in educational technology with a background in computer science will conclude after the first step that his or her sub-discipline is not represented in Mendeley and choose another one. 

\section{Conclusions and Future Work}
In this paper, we analyze the adequacy and applicability of readership statistics recorded in social reference management systems for creating knowledge
domain visualizations. We propose co-readership as a measure of subject similarity. An analysis of the distribution of subject areas in user libraries of educational technology researchers on Mendeley shows that 69.2\% of the journal articles in an average user library can be attributed to a single subject area. This is in line with an earlier study by \cite{Jiang2011} which finds that clusters based on the occurrence and co-occurrence of articles in user libraries of CiteULike are as effective as citation-based clusters. 

The prototypical visualization based on co-readership patterns of the field of educational technology comprises of 13 topic areas, which can be aggregated to meta-clusters, therefore strengthening the assumption that co-readership indicates subject similarity. The visualization is a recent representation of the field: 80\% of the publications included are from within ten years of data collection. However, not all of the latest developments were represented in the visualization due to the fact that it is harder to reach threshold values for the most recent publications. Nevertheless, the papers included in the co-readership analysis are on average almost half as young as the papers included in a comparable co-citation analysis by~\cite{Cho2012}. This suggests that co-readership analysis may be able to represent more recent aspects than co-citation. In order to generalize this statement and to better understand the differences between co-citation analysis, bibliographic coupling, and co-readership analysis, however, comparison studies between the different similarity measures must be carried out.

The characteristics of the readers introduce certain biases to the visualization. All scientometric analyses are subject to bias; it is therefore important that the characteristics of the underlying sample are made transparent. In the co-readership analysis, information encoded in the user profiles can be used to explain these characteristics. In the present study, a majority of readers were self-ascribed to the field of education and they came from an English-speaking country. This resulted in a map that represents an education science-dominated view from mainly an Anglo-American perspective.

%Biases affect all scientometric analyses. A problem that arises in citation studies is the selection of the corpus. Criteria for the inclusion of authors and papers in the analysis have an impact on the result. The difference between traditional citation-based analyses and the co-readership analysis is that in the latter case we do have more information about the users encoded in user profiles than is usually recorded about authors in bibliographic databases (e.g. career stage, education and professional experience).

One of the limitations of this work is that the methodology has only been tested for a single field of research. Educational technology is a diverse field with many influences; but it would not be appropriate to generalize the results to all research fields. The question is whether the same analysis would work as well on a larger set of publications and for other fields and disciplines. Each discipline has its own theories, methods, accepted practices, in short: its own culture. Just like publication and citation practices are fundamentally different for the natural sciences and the humanities, cultural differences might also affect the usage of social reference management systems. In the future, this study must therefore be repeated in other fields of research.This could be especially interesting for those fields that are dynamic in nature, and those that have not been scientometrically analyzed before due to the lack of citation data. 

When applied to larger collections of documents, the procedure used in this paper may be problematic. Both hierarchical clustering and multidimensional scaling have a high computational complexity. Therefore, it will be important to investigate algorithms that can deal with large data sets such as force-directed layout for ordination, and community detection for the establishment of topic areas. 
%Other highly scalable techniques of interest include singular value decomposition (SVD) and self-organizing maps (SOM). 
For a further discussion see \cite{Fortunato2010, Gibson2012}.

Finally, it seems promising to harness information encoded in the user profiles, such as location, discipline, and career stage, not only for a better understanding of the results (see above), but also for filtering the visualization. This would make it possible to compare visualizations, for instance between countries or career stages. Furthermore, with the availability of timestamps, it becomes possible to show the evolution of a research field over time at a granular level of detail.
%Data like this could be used to fuel pathfinder networks and other means of showing the development of a domain.

\section*{Acknowledgments}

We would like to thank the reviewers of this paper for their comprehensive comments and suggestions which have improved the paper considerably. The research presented in this work is in part funded by the European Commission as part of the FP7 Marie Curie IAPP project TEAM (grant no. 251514). The Know-Center is funded within the Austrian COMET program - Competence Centers for Excellent Technologies - under the auspices of the Austrian Federal Ministry of Transport, Innovation and Technology, the Austrian Federal Ministry of Economy, Family and Youth, and the State of Styria. COMET is managed by the Austrian Research Promotion Agency FFG.

\bibliography{references}

\end{document}